\documentclass{article}

\usepackage[preprint,nonatbib]{neurips_2020}

\usepackage[utf8]{inputenc} % allow utf-8 input
\usepackage[T1]{fontenc} 
\usepackage{graphicx}
\usepackage{xcolor}
\usepackage{booktabs}
\usepackage{subfig}
\usepackage{hyperref}

\usepackage{makecell}

\title{Localization of Malaria Parasites and White Blood Cells in Thick Blood Smears}

\author{
  Rose Nakasi\\
  Makerere University\\
  \texttt{g.nakasirose@gmail.com} \\
     \And
   Aminah Zawedde \\
  Makerere University\\
  \texttt{sazawedde@gmail.com} \\
   \And
    Ernest Mwebaze \\
  Sunbird AI\\
  \texttt{emwebaze@sunbird.ai} \\
    \And
  Jeremy Francis Tusubira \\
  Makerere University \\
  \texttt{tusubirafrancisjeremy@gmail.com} \\
   \And
  Gilbert Maiga \\
  Makerere University \\
  \texttt{gilmaiga@gmail.com} \\   
}

\begin{document}

\maketitle

\begin{abstract}
  Effectively determining malaria parasitemia is a critical aspect in assisting clinicians to accurately determine the severity of the disease and provide quality treatment. Microscopy applied to thick smear blood smears is the \emph{de facto} method for malaria parasitemia determination. However, manual quantification of parasitemia is time consuming, laborious and requires considerable trained expertise which is particularly inadequate in highly endemic and low resourced areas. This study presents an end-to-end approach for localisation and count of malaria parasites and white blood cells (WBCs) which aid in the effective determination of parasitemia; the quantitative content of parasites in the blood. On a dataset of slices of images of thick blood smears, we build models to analyse the obtained digital images. To improve model performance due to the limited size of the  dataset, data augmentation was applied. Our preliminary results show that our deep learning approach reliably detects and returns a count of malaria parasites and WBCs with a high precision and recall. We also evaluate our system against human experts and results indicate a strong correlation between our deep learning model counts and the manual expert counts (p=0.998 for parasites, p=0.987 for WBCs). This approach could potentially be applied to support malaria parasitemia determination especially in settings that lack sufficient Microscopists.

\end{abstract}

\section{Introduction}
\label{sec:intro}

\noindent Accurately determining malaria parasitemia, the quantitative content of malaria parasites in the blood, is key in determining measure of response to treatment, likely cause of disease and severity phases of malaria especially in non-immune patients \cite{WHO} \cite{Omeara2005}. However, most highly endemic  regions lack  sufficient skilled experts to efficiently determine parasitemia density of patients and  their manual counts of parasites and WBCs  are time consuming and prone to subjectivity and errors \cite{Killian2000}.

Different methods have been implemented to detect and count malaria parasites (trophozoites) and blood cells using image analysis techniques, but most have majorly been applied on thin blood smears \cite{Proudfoot2007} \cite{Sio2007} \cite{Le2008} \cite{Savkare2011} \cite{Poostchi2018}. 
According to WHO, thick blood smears are the gold standard for the determination of malaria parasitemia \cite{WHO2006}. Thick blood smears allow the microscopist to examine a larger number of red cells for the presence of parasites, and low parasitemia can more readily be identified in thick blood smears as compared to thin films \cite{Bejon2006}. Some previous work has used a semi-automated method for parasitemia determination in thick smears by using an open source java based image processing program \emph{ImageJ} (version 1.41) that is based on hand engineered features used to do manual counts of malaria parasites only \cite{Frean2009}.

 This study explores an end-to-end machine learning based algorithm for detection and counting of trophozoites and WBCs for improved parasitemia determination in thick blood smears. The pre-trained deep learning approach utilised is significantly faster, efficient and requires no additional operator preprocessing of the images for automated detection and counts. This is well suited for improving malaria parasitemia quantification in resource-constrained settings with few skilled lab technologists to interpret microscopy test results.

 \section{Methods}

\subsection{Data acquisition and preparation}
\label{dataacquisition}
Data was collected from a national referral hospital and annotated by malaria experts. Annotations are in form of bounding boxes around the parasites in an image of a blood slide. Proper permissions were obtained from both the patients and the hospital for this data collection.

A standardized data collection protocol was followed for this data collection. Thick blood slides where the experts had determined the presence of malaria parasites were used to generate the training data. From each blood slide, the microscope was adjusted to obtain different view point images. Each image with the corresponding set-up parameters including the microscope slide number, stage micrometer grid readings(x and y), zoom level of the smart phone used for image collection, objective size of the microscope, and the staining reagent used were captured.
Data collected from the lab was sent to a centralized server using an ODK form on a smartphone device.

To collect the data in the lab, a 3-D printed microscope attachment mechanism \cite{Quinn2016} was attached on an Olympus microscope with objective magnification (1000×). A Samsung J6 model smart phone with an installed phone zoom at 10 was attached onto the mounted 3-D printed device. Images captured with the phone and uploaded to the server, were downloaded and manual data annotations were performed by microscopists using the annotation tool LabelImg \cite{Tzutalin2015}. Annotations were saved in the Pascal VOC format \cite{Everingham2015}. The xml annotation file contained coordinates of bounding boxes of trophozoites and WBCs creating a multi-class object detection task. 
The annotated images were randomly split into a 9:1 train and test set. The dataset was encoded as a \emph{TFRecord} record; the data input format for a training job optimized for utilising the Tensorflow Object detection API.

\subsection{Model training and deployment }
\label{modeltraining}
In this study, we explored the potential of Faster R-CNN \cite{Girshick2015}  for the task of localization, and counting of trophozoites and WBCs in thick blood smears.  The model is an improvement to the Region based CNN model (R-CNN) whose detection procedure first uses Region Proposal Networks (RPNs) to extract features using selective search purposefully to generate proposals (boxes) which are further cropped and processed. For our case, Faster R-CNN was implemented with the fully convolutional Residual Network (ResNet 101) \cite{He2016}  as the backbone network. Comparison in performance between our proposed approach with single-class object detections (trophozoites-only) and (WBCs-only) was also conducted. Further, we compared our proposed model with a baseline Single Shot Multi Box Detector (SSD) \cite{Liu2016} for model performance comparison.

Training and testing was implemented on a 5$^{th}$ Gen Intel core i7 series processor with 16GB RAM and NVIDIA GTX 1060 GPU.
Models pre-trained on the COCO dataset \cite{Lin2014} were adopted in a transfer learning scheme to reduce training time. 
Data augmentation using random vertical and horizontal flipping of images was used to improve sample training size. An image size of 750 $\times$ 750 was configured as input to the Faster R-CNN model in a batch size of 1 (due to our low resources) with a base learning rate kept very low at 0.0003. For the baseline model, SSD MobileNet, images of 300x300 input size with a batch size of 12 at a learning rate of 0.0004 were considered for a light model like SSD MobileNet. 

Initially, the learning rate was kept at 0.0003 and 0.0004. However beyond 90,000 steps for each model, the learning rate was decreased to 0.00003 and 0.00004 for Faster R-CNN and SSD respectively. A momentum optimizer of 0.9 at 0.5 IOU threshold was maintained. Though 200,000 time steps were specified, each model attained optimal performance at different numbers of steps (see Table \ref{tab1}). Checkpoint files were frozen into a protobuff file to generate a tensorflow inference model. 

\section{Results and Discussion}
\label{results}
All models used were applied on a validation set of 100 images. An example of  model inference detections are shown in Figure [a]. It was observed that our proposed model performs and matches well with the ground truth as shown in Figure [b].
Results using the common metrics of mAP (mean Average precision) \cite{Cartucho2019}, precision and recall were used and results are summarised [see Table \ref{tab1}]. 

\begin{table*}[!h]
\centering
\caption{mAP@0.5, precision and recall performance for Faster R-CNN model for detection of malaria parasites and WBC.}
\begin{tabular}{@{}lcccc@{}}
\toprule
\multicolumn{1}{c}{\textbf{Algorithm}} & \textbf{Steps} & \textbf{mAP@0.5}  & \textbf{Precision} & \textbf{Recall} \\ \midrule
\thead{Faster R-CNN\\(trophozoites\_only)}      & 138,700         & 0.5506  & 0.6720             & 0.802                               \\
\thead{Faster R-CNN\\(WBC\_only)}    & 42,800         & 0.892  & 0.855             & 0.959                                          \\
\thead{Faster R-CNN\\(trophozoites + WBC)}                & 98,700          & 0.6609 & \thead{parasites: 0.686 \\ WBC: 0.805}    & \thead{parasites:0.9303 \\ WBC: 0.884}                    \\
\thead{SSD MobileNet\\(trophozoites + WBC)}                & 80,000          & 0.6292 & \thead{parasites: 0.760 \\ WBC: 0.89}    & \thead{parasites:0.501 \\ WBC: 0.806}                    \\
\bottomrule
\end{tabular}
\label{tab1}
\end{table*}

As shown in Table \ref{tab1}, Faster R-CNN performs better than the SSD MobileNet baseline model in terms of mAP and recall except in terms of precision. This model shows the importance and potential of Regional Proposal Networks for object localisation. 

Furthermore, results show that a multi-class detection task of both malaria trophozoites produces better results than that from single-class tasks  (trophozoites (only)), but lower than of a single-class of WBCs (0.892).  This could be due to the fact that WBCs are bigger objects and as such are easier examples to learn as compared to smaller objects (trophozoites only). The better performance of the multi-class task could be attributed to having bigger objects complementing small objects (trophozoites). Generally, the multi-class mAP value of 0.6609 (@0.5IOU) was considered to be good enough performance given the presence of smaller objects(trophozoites).

\begin{figure}[htbp]
\caption{Model counts Vs manual expert counts per film }
\centering
\begin{minipage}{0.7\textwidth}
\centering
\includegraphics[width=1.1\textwidth]{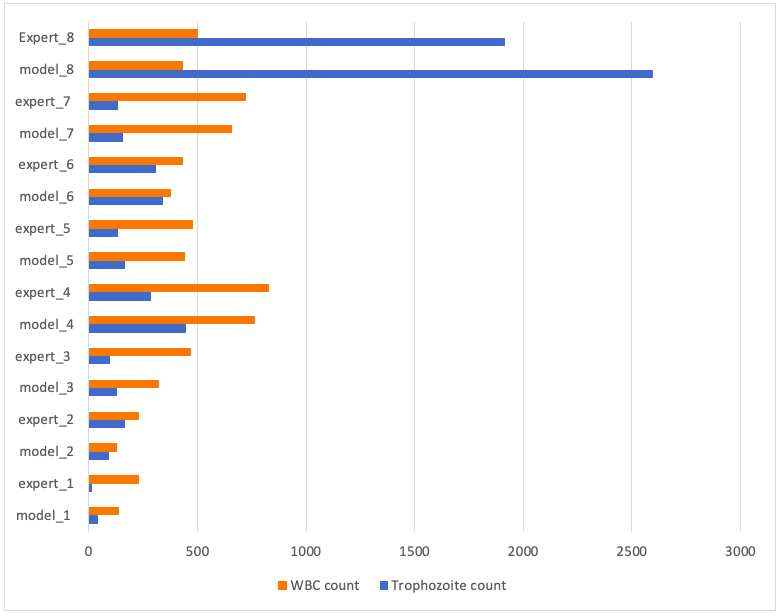}
\label{fig:Counts}
\end{minipage}%

\end{figure}

Furthermore, we show that our model-generated counts generally correlate well with manual-experts on a held-out dataset of 8 thick blood smear films with more than 100 images each (high Spearman's correlation coefficients for trophozoites (p=0.998) and WBCs (p=0.987)). In Figure \ref{fig:Counts}, model-generated parasite counts are slightly higher than expert-manual counts, this implies effect of false-positives that could be due to low quality annotations thus a lower precision registered for parasites.
In general the precision and recall values from our proposed model is high compared to the baseline model results.

\begin{figure}[!ht]
	\centering
	\subfloat[Groundtruth]{
		\includegraphics[width=.4\textwidth]{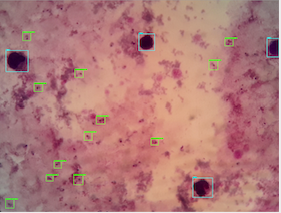}}
	\subfloat[Faster R-CNN]{
		\includegraphics[width=.4\textwidth]{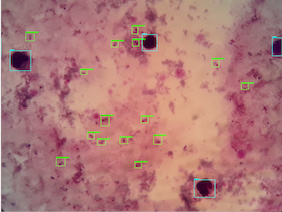}}

	\caption{Ground truth annotations (a) by lab experts Vs inference by Faster-RCNN model (b). The blue rectangles indicate WBCs and the green rectangles indicate malaria trophozoites.}
	\label{imageresults}
\end{figure}

Following from Table \ref{tab4}, film-wise Parasitemia density based on Parasites/$\mu$l method per film \cite{WHO2009}\cite{Kloub2019} was evaluated from the equation;

\begin{math}
    Parasites/\mathrm{\mu l}= \frac{\textrm{ No. of counted parasites} \times \textrm{No. of counted WBC}} {\textrm{No. of assumed WBC(8000)} }\\
\end{math}

\begin{table}[!h]
\centering
\caption{Malaria parasitemia on a unique set of each validation thick blood smear slide with corresponding clinical interpretation. }
\begin{tabular}{@{}lcccccccc@{}}
\toprule
\multicolumn{1}{c}{\textbf{Slide}}   & \textbf{1}&\textbf{2}&\textbf{3}&\textbf{4}&\textbf{5}&\textbf{6}&\textbf{7}&\textbf{8}   \\ \midrule
\textbf{\thead{Model\\Parasites/$\mu$l}}& 2326 & 5740 & 3160 & 4682  & 3011 & 7238  & 1906 & 47982  \\
\bottomrule
\end{tabular}
\label{tab4}
\end{table}

Malaria parasitemia clinical interpretations \cite{Garcia2015} were deduced, Slides 1,2,3,4,5,6 and 7 represent parasitemia level above which immune patients will exhibit malaria symptoms whereas Slide 8 represents a patient with maximum parasitemia.

\section{Conclusion and future work}
In this study, we provide an end-to-end machine learning approach for the determination of malaria parasite density on thick blood smears. We address the diagnostic challenges associated with conventional microscopy for parasitemia determination. Specifically, we leverage transfer learning and propose a Faster R-CNN ResNet 101 model pre-trained on COCO dataset for the task of object detection. We show that our proposed model can accomplish parasite and WBC localization and count effectively in a fast, accurate and consistent way for malaria parasitemia determination. In future, we intend to work towards reducing false positives and false negatives generated by the model. This could be through enforcing quality data annotation.

% \section*{References}
% \small
\bibliographystyle{plain}
%\bibliography{sample}
\bibliography{ml4d}

\end{document}